\documentclass[final,5p,times,fleqn]{elsarticle}
\usepackage{graphicx}
\usepackage{amssymb}
\usepackage{mathrsfs}
\usepackage{amsmath}
\usepackage{comment}
\usepackage{amsbsy}
\usepackage{mathrsfs}
\usepackage{amsfonts}
\usepackage{lineno}
\usepackage{nicefrac}
\usepackage{epsf,color,colordvi,pifont}
\usepackage[notref,notcite]{showkeys} 
\biboptions{sort&compress}
\def\deltabar{{\mathchar'26\mkern-8mu\delta}}
\def\dbar{{\mathchar'26\mkern-12mu d}}
\DeclareFontFamily{OT1}{pzc}{}
\DeclareFontShape{OT1}{pzc}{m}{it}{<-> s * [1.10] pzcmi7t}{}
\DeclareMathAlphabet{\mathpzc}{OT1}{pzc}{m}{it}

\def\dbar{{\mathchar'26\mkern-12mu d}}
\begin{document}

\begin{frontmatter}
\title{Polarization-operator approach to optical signatures of  axion-like particles\\  in strong laser pulses}
\author{S. Villalba-Ch\'avez}
\ead{selym@tp1.uni-duesseldorf.de}
\author{T. Podszus}
\ead{tobias.podszus@hhu.de}
\author{C. M\"{u}ller}
\ead{c.mueller@tp1.uni-duesseldorf.de}
\address{Institut f\"{u}r Theoretische Physik I, Heinrich-Heine-Universit\"{a}t D\"{u}sseldorf, Universit\"{a}tsstr. 1, 40225 D\"{u}sseldorf, Germany}
\date{\today}

\begin{abstract}
Hypothetical oscillations  of probe photons into axion-like particles might be revealed by exploiting the  strong  fields of high-intensity  laser pulses.
Considering an arbitrary plane-wave background, we determine the polarization tensor induced by the quantum fluctuations of the  axion field and use it to calculate 
how  the polarimetric  properties of an initially linear-polarized probe  beam are modified. We  find  that  various  experimental 
setups based on contemporary facilities and instrumentation might lead to new exclusion  bounds  on the  parameter space of these  particle  candidates. The impact 
of the  pulse shape on the  discovery  potential is  studied  via a comparison between the cases in which the wave is modulated by a Gaussian envelope and a 
$\sin^2$ profile. This analysis shows that the upper limits  resulting from  the ellipticity are relatively  insensitive to this change, whereas those arising from  
the  rotation of the polarization plane  turn out to be more dependent on the field shape.
\end{abstract}
\begin{keyword}
Beyond the Standard Model\sep Axion-like particles\sep Vacuum polarization\sep Laser Fields.
\PACS 14.80.-j \sep 12.20.Fv 
\end{keyword}
\end{frontmatter}

\section{Introduction}

A spontaneous breakdown of the global $\rm U(1)-$Peccei-Quinn symmetry occurs in the  course of explaining the absence of charge-parity violation in the 
theory of  strong interactions. The emerging Nambu-Goldstone boson--known as the QCD axion \cite{Peccei:1977hh,Wilczek:1977pj,Weinberg:1977ma}--constitutes  
the flag representative for the axion  models \cite{Dine:1981rt,zhitnitskii,kim,shifman} and  the class of  Axion-Like Particles (ALPs) that are 
predicted in conformal scenarios \cite{Meissner:2007xv},  string  theory \cite{Witten:1984dg,Svrcek:2006yi,Lebedev:2009ag,LCicoli:2012sz} as well as  various Standard 
Model extensions, where they are linked to  dark  matter \cite{covi,Raffelt:2006rj,Duffy:2009ig,Sikivie:2009fv,Baer:2010wm}. Despite experimental efforts 
toward their detection--much of them exploiting their coherent oscillations into photons mediated by a static magnetic field--there is no evidence yet of  
ALPs.  This  fact  manifests  that their  interactions with the well established Standard Model  branch 
might be extremely weak,  and the absence of  positive detection signals can be used to constrict  the associated  parameter space. Stringent  
upper bounds on  the ALP-diphoton coupling $g$  have  been inferred  from astrophysical considerations. A plausible generation of ALPs in the core  
of stars via the Primakoff process might lead to  an energy  loss which  accelerates their  cooling  and, therefore, their lifetimes.  The nonobservation 
of a diminishing in the number of  stars in the Helium-burning phase [horizontal-branch (HB) stars] within  globular clusters  constraints  $g$ to lie below  $g\lesssim 10^{-10}\ \rm GeV^{-1}$ 
for  ALP masses $m$ below the $\rm keV$ scale   \cite{Raffelt:1985nk,Raffelt:1999tx,Raffelt:2006cw}. Furthermore, as these particle candidates  may  escape from the Sun  almost freely,  
solar ALPs  would  likely arrive to  Earth.   By  monitoring these hypothetical ALP fluxes, the CAST collaboration has obtained  $g<9\times 10^{-11}\ \rm GeV^{-1}$, 
whenever   $m$ is below $\rm 10\  meV$  \cite{CAST2011,CAST2014}.  

Bounds resulting from laboratory  experiments are considerably less stringent but  free from the uncertainty  associated with the underlying 
astrophysical models  \cite{Masso:2005ym,Masso:2006gc,Jaeckel:2006id,evading}.  Some of them  have been established from the search of Light Shining through a 
Wall (LSW) \cite{Chou:2007zzc,Afanasev:2008jt,Steffen:2009sc,Pugnat:2007nu,Robilliard:2007bq,Fouche:2008jk,Ehret:2010mh,Balou} and from  scenarios oriented to detect  
the  magnetically-induced  vacuum  dichroism   and  birefringence mediated by real and virtual ALPs, respectively  \cite{Cameron:1993mr,PVLAS2008,BMVreport,Chen:2006cd,Mei:2010aq,DellaValle:2013xs}. 
While in LSW  setups,  the best  upper limit is held by the OSCAR collaboration $g<4\times 10^{-8}\ \rm GeV^{-1}$ \cite{Balou},   the current  best bound resulting from  
polarimetric studies has been  established  by PVLAS  $g<8\times 10^{-8}\ \rm GeV^{-1}$  \cite{DellaValle:2013xs}. These limits apply for $m\lesssim \rm 100\  \mu eV$,  
and  relax significantly  for masses larger than $m>10 \ \rm meV$ by several orders of magnitude.  As a general  feature both, LSW and polarimetric setups, might  improve their  
respective bounds  by increasing  the  field strength   and the  distance over which it   extends.  At present, the 
largest magnetic field generated  by  superconducting  dipole  magnets amounts to $\sim10^6\ \mathrm{G}$  over  a length  smaller than $10\ \rm m$. The incorporation  
of interferometric cavities for the probe  beam allows for extending the interaction region upto five orders of magnitude, but its use has  not been  enough to push 
down the bounds in regions of  masses  $m>10\  \rm meV$,  where they turn out to be much less stringent. 

Higher field strengths $\sim 10^9\,{\rm G}$ can be obtained nowadays within  the focal spots of high-intensity laser pulses.  Even larger magnetic fields   $\sim 10^{11}\ {\rm G}$, i.e. two orders
of magnitude below the critical scale $B_c=4.42\times 10^{13}\ \rm G$ of Quantum Electrodynamics (QED), are  envisaged in the near future within the  ELI and XCELS projects \cite{ELI,xcels}. Despite the 
inhomogeneous nature of these pulses--confined to  short spatial extensions $\sim \text{$\mu$m}$--their use may allow for the realization of various  elusive QED  processes \cite{Di_Piazza_2012}. For instance, the   
HIBEF collaboration  \cite{HIBEF,Schlenvoigt} has  put  forward a laser-based experiment  with which vacuum  birefringence \cite{Heinzl,Dinu:2013gaa,Dinu:2014tsa,King:20161}  
may soon  be detected. This kind of experiment provides propitious arenas to test the frontier of the Standard Model at low energies \cite{Jaeckel:2010ni,Ringwald:2012hr,Hewett:2012ns,Essig:2013lka} and therefore, 
complements those setups driven by particle accelerators. Various theoretical studies in this direction have been  carried out to estimate  whether  high-intensity laser pulses are feasible in the searches for ALPs 
\cite{mendonza,Gies:2008wv,Dobrich:2010hi,Tommasini:2009nh,Villalba-Chavez:2013bda,Villalba-Chavez:2013goa}, minicharged particles and hidden photons \cite{Villalba-Chavez:2013gma,Villalba-Chavez:2014nya,Villalba-Chavez:2015xna,Villalba-Chavez:2016hht}. 
However, the complicated nature of pulsed  laser fields  makes  the phenomenological descriptions  rather challenging and  full   characterizations of these problems  are
far from  being  complete. 

In  this Letter, we provide a  step toward the understanding  of the role  that  the pulse profile may play in the search for ALPs, this way  extending  a recent investigation carried out within the context of  minicharged 
particles \cite{Villalba-Chavez:2016hht}.   We show that, when dealing with   a  polarimetric probe  driven by the field of a high-intensity linearly polarized  pulse, the upper limits  resulting from  the ellipticity are 
almost insensitive to the pulse shape, whereas those  arising from  the  rotation of the polarization plane  turn out to be more  dependent on the field profile. Besides, we reveal that  this kind of setup might 
notably  improve the existing  laboratory  limits in some regions of the ALP parameter space.   Our investigation relies on the polarization  tensor induced by the quantum fluctuations of the axion field over a  
plane-wave background. 

\section{Photon propagation in the  vacuum of ALPs \label{GP}}

\begin{figure}
\begin{center}
\includegraphics[width=7cm]{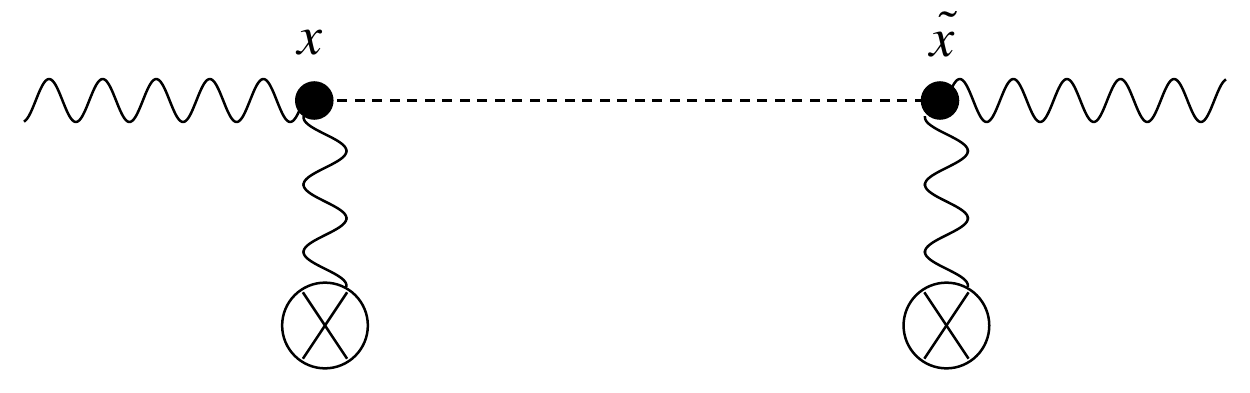}
\caption{\label{fig.001} Diagrammatic representation   of the vacuum polarization tensor  mediated by a quantum fluctuation of a pseudoscalar field $\phi(x)$ in a high-intensity laser pulse 
[vertical wavy lines].  Here,  the  dashed line represents  the  ALP propagator $\Delta_\mathrm{F}(x,\tilde{x})$, whereas the horizontal wavy lines must be understood as the amputated photon legs. }
\end{center}
\end{figure}

Searches for axion  dark matter   rely  on the existence of an ALPs background  permeating the universe. 
In the following we  will assume  that the effects resulting from this nontrivial  expectation value are negligible in 
comparison with the quantum fluctuations that are induced by a pseudoscalar field $\phi(x)$  on the propagation 
of  a small-amplitude electromagnetic wave $a_\mu(x)$.  We are interested in evaluating  these  effects in an external 
electromagnetic field characterized by the tensor $\mathscr{F}^{\mu\nu}=\partial^\mu \mathscr{A}^\nu-\partial^\nu \mathscr{A}^\mu$  
with $\mathscr{A}_{\mu}(x)$ denoting its  four-potential. As long as  the fields of interest  are minimally 
coupled, preserving  the formal invariance properties of QED, the relevant equations of motion are\footnote{From now on ``natural'' 
and Gaussian  units  $c=\hbar=4\pi\epsilon_0=1$ are used. Besides, the  metric tensor $\mathpzc{g}_{\mu\nu}$ 
is taken with signature $(+1,-1,-1,-1)$  so that  $\mathpzc{A}\mathpzc{B}=\mathpzc{A}_0\mathpzc{B}_0-\pmb{\mathpzc{A}}\cdot\pmb{\mathpzc{B}}$.} 
\begin{eqnarray}
\begin{split}
&\square a^\mu(x)+g\tilde{\mathscr{F}}^{\mu\nu}(x)\partial_\nu\phi(x)=0,\\  
&\left(\square+m^2\right)\phi(x)-\frac{1}{8\pi}g\tilde{\mathscr{F}}^{\mu\nu}(x)f_{\mu\nu}(x)=0,
\end{split}
\label{DSE}
\end{eqnarray} provided  that  $a_\mu(x)$  is   chosen in the Lorenz gauge $\partial_\mu a^\mu =0$ and $\square\equiv\partial_\mu\partial^\mu=\partial^2/\partial t^2-\nabla^2$.  
Here the  coupling  constant $g$ and  mass $m$  are unknown parameters, $f^{\mu\nu}=\partial^\mu a^\nu-\partial^\nu a^\mu$,  whereas  the dual of the external field tensor is 
$\tilde{\mathscr{F}}^{\mu\nu}=\frac{1}{2}\epsilon^{\mu\nu\alpha\beta}\mathscr{F}_{\alpha\beta}$. When solving the second  equation  involved in (\ref{DSE}) and 
substituting the expression for $\phi(x)=\frac{g}{8\pi(\square+m^2)}\tilde{\mathscr{F}}^{\mu\nu}(x)f_{\mu\nu}(x)$   into the equation associated with the small-amplitude wave, we end up with  
\begin{equation}
\begin{split}
&\square a^\mu(x)+\int d^4\tilde{x}\ \Pi^{\mu\nu}(x,\tilde{x})a_\nu(\tilde{x})=0,\\
&\Pi^{\mu\nu}(x,\tilde{x})=-\frac{ig^2}{4\pi} \tilde{\mathscr{F}}^{\mu\tau}(x)\left[\partial_\tau^{x}\partial_\sigma^{\tilde{x}}\Delta_{\mathrm{F}}(x-\tilde{x})\right]\tilde{\mathscr{F}}^{\sigma\nu}(\tilde{x}),
\end{split}\label{PHOTONEQMOTION}
\end{equation}where  $\Delta_{\mathrm{F}}(x-\tilde{x})=\int \frac{d^4p}{(2\pi)^4}\frac{i}{p^2-m^2+i0}e^{-ip(x-\tilde{x})}$ denotes the  ALP propagator. Note that Eq.~(\ref{PHOTONEQMOTION}) has been written in a way that   
resembles  the effective equation of  motion of the electromagnetic field  in QED, i.e., including the photon radiative correction. This fact  allows us to identify 
straightforwardly  the  polarization tensor $\Pi^{\mu\nu}(x,\tilde{x})$ induced by the quantum vacuum  fluctuations of the pseudoscalar field $\phi(x)$. 
The Feynman diagram associated with this tensor is   depicted in  Fig.~\ref{fig.001}.

Hereafter, we indentify the background described by $\mathscr{F}^{\mu\nu}(x)$  with the field produced by a high-intensity laser, the  potential of which  is taken in  the form 
\begin{equation}\label{extfield}
\mathscr{A}^{\mu}(x)=\mathpzc{a}^{\mu}_{1}\psi_1(\varphi)+\mathpzc{a}^{\mu}_{2}\psi_2(\varphi),
\end{equation} where $\mathpzc{a}_{1,2}$ are two orthogonal amplitude vectors [$\mathpzc{a}_{1}\mathpzc{a}_{2}=0$]  and   $\psi_{1,2}(\varphi)$  arbitrary  functions of  $\varphi=\varkappa x$   satisfying  
the  boundary conditions  $\psi_{1,2}(\pm\infty)=\psi_{1,2}^\prime(\pm\infty)=0$.    Here, the prime denotes the derivative with respect  to the argument, $\psi_{1,2}^\prime(\varphi)\equiv d\psi_{1,2}(\varphi)/d\varphi$.  The external potential is chosen  in  the Lorenz  gauge $\partial_{\mu}  \mathscr{A}^\mu=0$  so that   the wave four-vector  $\varkappa^{\mu}=(\varkappa^0,\pmb{\varkappa})$  
with $\varkappa^2=0$  and  the   amplitude  vectors  $\mathpzc{a}_{1,2}^\mu$   satisfy the constraints  $\varkappa \mathpzc{a}_{1,2}=0$.  The presence of this plane-wave background allows for  introducing   
the   four-vectors
\begin{eqnarray}\label{vectorialbasisbaier}
\Lambda_{1,2}^\mu(q)=-\frac{\mathscr{F}_{1,2}^{\mu\nu}q_\nu}{q\varkappa\sqrt{-\mathpzc{a}_{1,2}^2}},\quad \tilde{\Lambda}_{1,2}^\mu(q)=-\frac{\Tilde{\mathscr{F}}_{1,2}^{\mu\nu}q_\nu}{q\varkappa\sqrt{-\mathpzc{a}_{1,2}^2}},
\end{eqnarray} which are built up from  the amplitudes of the  external  field modes  $\mathscr{F}^{\mu\nu}_{i}=\varkappa^\mu\mathpzc{a}^\nu_{i}-\varkappa^\nu\mathpzc{a}^\mu_{i}$ [$i=1,2$]. These  vectors are transverse  
$q_\mu\Lambda_{1,2}^\mu(q)=q_\mu\tilde{\Lambda}_{1,2}^\mu(q)=0$, orthonormalized according to    $\Lambda_{i}^\mu(q)\Lambda_{j\mu}(q)=\tilde{\Lambda}_{i}^\mu(q)\tilde{\Lambda}_{j\mu}(q)=-\delta_{ij}$ 
and satisfy the relation $\tilde{\Lambda}_i^\mu(q)\Lambda_{j\mu}(q)=-\epsilon_{ij}$ with   $i,j=1,2$. Here the antisymmetric tensor $\epsilon_{ij}$ with  $\epsilon_{12}=1$ is used. 

In the following, we  Fourier transform  Eq.~(\ref{PHOTONEQMOTION}) and  seek the solutions  of the resulting  equation of motion in the form of a superposition of transverse waves 
$a^\mu(q)=\sum_{i=1,2}\Lambda^\mu_{i}(q)f_i(q)$. Correspondingly, 
\begin{equation}
\begin{split}
&q_2^2f_i(q_2)=-\sum_j\int \frac{d^4 q_1}{(2\pi)^4} \Lambda_j^\mu(q_1)\Pi_{\mu\nu}(q_1,q_2)\Lambda^{\nu}_i(q_2)f_j(q_1),\\
&\Pi_{\mu\nu}(q_1,q_2)=\int d^4x\ d^4\tilde{x}\ e^{-iq_1x} \Pi_{\mu\nu}(x,\tilde{x})e^{iq_2\tilde{x}}.
\end{split}\label{fourier}
\end{equation}In obtaining the expression above we have used the symmetry property $\Pi_{\mu\nu}(-q_2,-q_1)=\Pi_{\nu\mu}(q_1,q_2)$.
From now on,  we  choose  the  reference frame in such a way  that the direction of  propagation of our  external plane wave  [see Eq.~(\ref{extfield})] is along the positive direction of the  third axis. 
As  a consequence,   the external field only   depends on  $x_-=(x^0-x^3)/\sqrt{2}$ via $\varphi=\varkappa_+x_-$  with  $\varkappa_+=(\varkappa^0+\varkappa^3)/\sqrt{2}=\sqrt{2}\varkappa_0>0$.  

Next, we insert the polarization  tensor in position space [see Eq.~(\ref{PHOTONEQMOTION})] into Eq.~(\ref{fourier}). Afterwards, integrations  by parts over $x$ and $\tilde{x}$ are 
carried out considering the boundary condition  $\psi_{1,2}^\prime(\pm \infty)=0$. Later, we introduce the  light-cone  variables  $x_\pm=(x^0\pm x^3)/\sqrt{2}$, $\pmb{x}_\perp=(x^1,x^2)$ and $\tilde{x}_\pm=(\tilde{x}^0\pm \tilde{x}^3)/\sqrt{2}$, $\tilde{\pmb{x}}_\perp=(\tilde{x}^1,\tilde{x}^2)$. 
Their use allows us to integrate six out of the eight variables involved in $\Pi^{\mu\nu}(q_1,q_2)$. As a consequence,
\begin{equation}
\begin{split}
&\Pi^{\mu\nu}(q_1,q_2)=\frac{\delta_{q_2,q_1}}{\varkappa_+}\int d\tilde{\varphi}\  \mathpzc{P}^{\mu\nu}(\tilde{\varphi},q_1,q_2) e^{\frac{iq_{2+}}{\varkappa_+}\tilde{\varphi}}, 
\end{split}
\end{equation}where the notation $\delta_{q_2,q_1}=(2\pi)^3\delta^{(\perp)}(q_2-q_1)\delta^{(-)}(q_2-q_1)$ has been introduced.  The tensorial structure  of 
$\mathpzc{P}^{\mu\nu}(\tilde{\varphi},q_1,q_2)$ resembles the one associated with the polarization tensor of  QED \cite{Villalba-Chavez:2016hht}: 
\begin{equation}
\begin{split}
&\mathpzc{P}^{\mu\nu}(\tilde{\varphi},q_1,q_2)=c_1\Lambda^\mu_{1}(q_1)\Lambda^\nu_{2}(q_2)+c_2\Lambda^\mu_{2}(q_1)\Lambda^\nu_{1}(q_2)\\ &\qquad\qquad\qquad+c_3\Lambda^\mu_{1}(q_1)\Lambda^\nu_{1}(q_2)+c_4\Lambda^\mu_{2}(q_1)\Lambda^\nu_{2}(q_2).
\end{split}\label{DVPBaier}
\end{equation}As  $q_1-q_2\sim \varkappa$,  this   decomposition  does not depend on which choice of $q$ is taken; see also Eq.~(\ref{vectorialbasisbaier}). The involved form factors  $c_{i}$  depend on the phase of 
the external field $\tilde{\varphi}$, $q_1$ and $q_2$.  Explicitly,
\begin{equation}
\begin{split}
&c_1=\frac{g^2}{2\varkappa_0^2}(\varkappa q_{2})\sqrt{I_1I_2}\int\frac{d\eta}{2\pi}\frac{\tilde{\psi}_2(\eta_{q_1})\psi_1^\prime(\tilde{\varphi})}{\eta-\frac{q_{1\perp}^2+m^2}{2\varkappa q_{1}}+i0} e^{-i\eta\tilde{\varphi}},\\
&c_3=-\frac{g^2}{2\varkappa_0^2}(\varkappa q_{2})I_2\int\frac{d\eta}{2\pi}\frac{\tilde{\psi}_2 (\eta_{q_1})\psi_2^\prime(\tilde{\varphi})}{\eta-\frac{q_{1\perp}^2+m^2}{2\varkappa q_{1}}+i0} e^{-i\eta\tilde{\varphi}},\\
&c_2=c_1(1\leftrightarrow2),\quad c_4=c_3(1\leftrightarrow2),
\end{split}\label{formfactors}
\end{equation}where $\eta_{q_1}\equiv\eta-(q_{1}^2+q_{1\perp}^2)/(2 \varkappa q_1)$ and the change of variable $\eta=-p_+/\varkappa_+$ has been carried out. In  Eq.~(\ref{formfactors}), the exchange 
$1\leftrightarrow2$ must be  carried out only on the index of the   field profile functions $\psi^\prime_{1,2}$ and on the  peak intensity associated with each external field mode  $I_{1,2}=E_{1,2}^2/(4\pi)$ with $E_{1,2}^2=-\varkappa_0^2\mathpzc{a}_{1,2}^2$. Here  
$\tilde{\psi}_{1,2}(\alpha)=\int d\varphi\ \psi_{1,2}^\prime(\varphi)e^{i\alpha\varphi}$ is the Fourier transform of   $\psi_{1,2}^\prime(\varphi)$. We remark that also  other  representations  for  $c_i$  can be found. 
We will see, however,  that the chosen one turns out to be  convenient for the purposes of this work.

We  solve Eq.~(\ref{PHOTONEQMOTION})  by following a procedure similar to the one used   in the context of minicharged particles [see Ref.~\cite{Villalba-Chavez:2016hht}].
If the ALP effects do not  modify  the Maxwell equations dramatically,  one can  solve Eq.~(\ref{fourier}) perturbatively by setting 
$f_{i}(q)\approx f_{0i}(q)+\delta f_i(q)$. In the following, we suppose a head-on  collision between  the strong laser pulse and the probe beam characterized by  a four-momentum  
$k^\mu=(\omega_{\pmb{k}},\pmb{k})$,  so that $\varkappa_+k_{-}=2\omega_{\pmb{k}}\varkappa_0$ and $\pmb{k}_{\perp}=\pmb{0}$. 
In accordance, the leading order term is  $f_{0i}(q)=\vert 2q_-\vert a_{0 i}(2\pi)^4 \delta(q^2)\ \delta^{(\perp)}(q)\ \delta^{(-)}(q-k)$, corresponding to  $f_{0i}(x)=a_{0i}e^{-i\phi}$ 
with  $\phi=kx=k_{-}x_+$ and $a_{0i}$ the amplitude of mode-$i$. Besides, it follows from Eq.~(\ref{fourier}) that  the perturbative contribution is given by
\begin{equation}
\begin{split}
&\delta f_i(q_{2})=-\frac{\delta_{q_2,k}}{\varkappa_+\left[2q_{2+}q_{2-}-q_{2\perp}^2+i0\right]}\sum_ja_{0j}\int d\tilde{\varphi}\\ &\qquad\qquad\times\Lambda_{j}^\mu(k)\mathpzc{P}_{\mu\nu}(\tilde{\varphi},k,q_2) \Lambda^\nu_i(q_2)e^{\frac{iq_{2+}}{\varkappa_+}\tilde{\varphi}},
\end{split}
\end{equation}where it must be understood that the only nonvanishing light-cone component of the four-vector $k^\mu$   is $k_-$.  Besides, the poles in the function  $1/q_2^2$ have been shifted  
infinitesimally into the complex plane  by an $i0$-term  so that   correct boundary conditions  of the fields at asymptotic times  $f_i(\pm\infty,\pmb{x})$ are implemented. When Fourier transforming 
back, the solution  of our  problem  reads $a^\mu(x)=\sum_{i=1,2}\Lambda_i^\mu(k) f_i(x)$ [see above Eq.~(\ref{fourier})], where  
\begin{equation}\label{intermediateequation}
\begin{split}
&f_i(x)\approx e^{-i\phi}\left[a_{0i}-\frac{1}{2\varkappa_+k_-}\sum_{j}a_{0j}\int d\tilde{\varphi}\int\frac{dq_{2+}}{2\pi}\right.\\ &\qquad\qquad\qquad\quad\times\left.\Lambda_j^\mu(k)\frac{\mathpzc{P}_{\mu\nu}(\varphi,k,q_2)}{q_{2+}+i0}\Lambda^\nu_i(q_2)e^{\frac{iq_{2+}}{\varkappa_+}(\tilde{\varphi}-\varphi)}\right].
\end{split}
\end{equation}Here, $q_{2-}=k_-$, $\pmb{q}_{2\perp}=\pmb{0}$,  whereas  $\pmb{k}_\perp=\pmb{0}$ and $k_+=0$. We remark that, in our reference frame,  the  transversality  condition $\varkappa \mathpzc{a}_{1,2}=0$ 
[see below Eq.~(\ref{extfield})] implies  that  $\mathpzc{a}_{1,2-}=0$. It can be  verified that such a constraint implies $\Lambda^\mu_{1,2}(q)$ to be independent of $q_{+}$. This means that, in the expression above  
$\Lambda^\nu_i(q_2)=\Lambda^\nu_i(k)$. Structurally, Eq.~(\ref{intermediateequation}) coincides with Eq.~(8) found in 
Ref.~\cite{Villalba-Chavez:2016hht}. This fact allows us to   integrate  out  $q_{2+}$ by using the procedure  explained there. As a consequence of this assessment, the integration over  
$\tilde{\varphi}$  turns out to be  restricted to the kinematically allowed region  $(-\infty,\varphi]$ and we end up with 
\begin{equation}\label{degradedDSE}
\begin{split}
&f_i(x)\approx f_{0i}(x)\\ &\qquad+\frac{i}{2\varkappa_+k_-}\sum_{j}f_{0j}(x)\int_{-\infty}^\varphi  d\tilde{\varphi}\ \Lambda_j^\mu(k)\mathpzc{P}_{\mu\nu}(\varphi,k,k)\Lambda^\nu_i(k).
\end{split}
\end{equation} The expression above  constitutes the starting point for further considerations. It  holds for arbitrary plane-wave  profiles, which formally implies that the  pulsed field  is infinitely 
extended in the plane perpendicular to the propagation direction. This means that, in our model,  ALPs do not experience transverse focusing effects.   In an actual experimental realization,  this condition can be  
considered as  satisfied whenever the ALP Compton wavelength  $\lambda_{\mathrm{ALP}}=1/m$ turns out to be much smaller than the characteristic spatial scale,  set  by the  waist size  of the pulse $w_0$. 
In order words,  the outcomes resulting from Eq.~(\ref{degradedDSE}) are expected to be trustworthy for  ALP masses $m \gg w_0^{-1}$. 

When  the external plane wave [see Eq.~(\ref{extfield})] is   linearly polarized   with $\mathpzc{a}_2=0$, $\psi_2(\varphi)=0$, the probe modes in  Eq.~(\ref{degradedDSE}) disentangle from each other. Consequently, we 
can write  the  electric field of the probe   [$\pmb{\varepsilon}=-\partial\pmb{a}/\partial x^0$ with $a_0=0$] as a superposition of waves
\begin{eqnarray}
\begin{split}
&\pmb{\varepsilon}(x)\approx\varepsilon_0\cos(\vartheta_0)\pmb{\Lambda}_{1} \mathrm{Re}\ \exp\left[-i\phi\right]\\ 
&\qquad+\varepsilon_0\sin(\vartheta_0)\pmb{\Lambda}_{2}\mathrm{Re} \exp\left[-i\phi+\frac{i}{2\varkappa_+k_-}\int_{-\infty}^\varphi d\tilde{\varphi}\ c_4(\tilde{\varphi})\right]. \\
\end{split}\label{electricfield}
\end{eqnarray}where the approximation $1+ix \approx \exp(ix)$ has been used. In the expression above  only the leading   term,  which does not vanish at  asymptotically  large spacetime distances  [$\varphi\to\infty$],  
when the  high-intensity  laser  field  is turned off, has been considered. Here  $\varepsilon_0$ refers to the initial  electric 
field amplitude, $\pmb{\Lambda}_{1,2}=\pmb{\mathpzc{a}}_{1,2}/\vert\pmb{\mathpzc{a}}_{1,2}\vert$,  whereas  $0\leqslant\vartheta_0<\pi/2$ is the  corresponding  initial polarization  angle  of the 
probe with  respect to $\pmb{\Lambda}_{1}$, i.e., the polarization axis of the external pulse. In the expression above, the form factor $c_4(\tilde{\varphi})$ [see Eq.~(\ref{formfactors})] must 
be evaluated with $q_2=q_1=k$. Therefore, 
\begin{eqnarray}
\begin{split}
&c_4(\tilde{\varphi})=-\frac{g^2}{4\varkappa_0^2}(\varkappa_+ k_-)I_1\psi_1^\prime(\tilde{\varphi})\frac{1}{\pi}\mathrm{P}\int_{-\infty}^{\infty}d\eta \frac{\tilde{\psi}_1(\eta)e^{-i\eta\tilde{\varphi}}}{\eta-\mathpzc{n_*}}\\&\qquad+\frac{ig^2}{4\varkappa_0^2}(\varkappa_+ k_-)I_1\psi_1^\prime\left(\tilde{\varphi}\right)\tilde{\psi}_1\left(\mathpzc{n}_*\right)e^{-i\mathpzc{n}_*\tilde{\varphi}},
\end{split}\label{relevantformfactor}
\end{eqnarray}where  the relation $(x+i0)^{-1}=\mathrm{P}\frac{1}{x}-i\pi\delta(x)$, with  $\rm P$ refering to the Cauchy principal value,  has been  applied and  $\mathpzc{n}_*=m^2/(2\varkappa_+k_-)$ denotes the resonant parameter.

Besides, in this external field configuration, the total probability that a photon with 
polarization $\Lambda_2$ does not decay inside the laser  pulse is obtained by  evaluating  the square of the wave function, $
\mathcal{P}_{\gamma\to\gamma}(\varphi)=\vert\Lambda^\mu_{2}f_2(\phi,\varphi)\vert^2/\vert a_{02}\vert^2=1-\mathcal{P}_{\gamma\to\phi}(\varphi)$. Here,
\begin{equation}
\mathcal{P}_{\gamma\to\phi}(\varphi)=\frac{g^2I}{4\varkappa_0^2}\tilde{\psi}_1\left(\mathpzc{n}_*\right) \int_{-\infty}^\varphi d\tilde{\varphi}\ \psi_1^\prime\left(\tilde{\varphi}\right)e^{-i\mathpzc{n}_*\tilde{\varphi}}
\end{equation} refers to  the probability that a photon  oscillates into an ALP,  a phenomenon  which damps the intensity of the probe  beam $I(\varphi)=\frac{\varepsilon_0^2}{4\pi} \left[\cos^2(\vartheta_0)+ \sin^2(\vartheta_0) \exp(-\kappa)\right]$ 
as it propagates in the pulse. The  factor  responsible for the damping is  $\kappa(\varphi)\approx\mathcal{P}_{\gamma\to\phi}(\varphi)$, provided $\kappa(\varphi)\ll1$. Therefore,  the vacuum  behaves  like  a dichroic  medium, inducing a rotation of the probe polarization  from the initial angle  $\vartheta_0$ to $\vartheta_0+\delta\vartheta$, where  
$\delta \vartheta$ is expected to be tiny. At asymptotically  large spacetime distances  [$\varphi\to\infty$],  we find 
\begin{eqnarray}\label{rot}
\begin{split}
&\delta\vartheta(g,m)\approx-\frac{1}{4}\sin(2\vartheta_0)\mathcal{P}_{\gamma\to\phi}(\infty),\\ &\mathcal{P}_{\gamma\to\phi}(\infty)=\frac{g^2I}{4\varkappa_0^2}\vert\tilde{\psi}_1\left(\mathpzc{n}_*\right)\vert^2.
\end{split}
\end{eqnarray}
As the phase difference  between the two propagating modes, $\delta\phi(\varphi)=\frac{1}{2\varkappa_+k_-}\mathrm{Re}\ \int_{-\infty}^{\varphi} d\tilde{\varphi}\ c_4(\tilde{\varphi})$,  
does not vanish  either,  the vacuum is also predicted  to be  birefringent. Hence,  when the strong field is turned off [$\varphi\to\infty$],  
the  outgoing probe should be elliptically polarized and its  ellipticity   is given by \cite{Born}
\begin{equation}\label{elip}
\vert\psi(g,m)\vert\approx \frac{1}{2}\sin(2\vartheta_0)\frac{g^2I}{8\varkappa_0^2}\left\vert-\frac{1}{\pi} \mathrm{P}\int_{-\infty}^{\infty}d\eta \frac{\vert\tilde{\psi}_1(\eta)\vert^2}{\eta-\mathpzc{n}_*}\right\vert.
\end{equation}  We remark that  the last formula is a good approximation only when  its right-hand side   is smaller than unity. Manisfestly, the relation between Eq.~(\ref{rot}) and (\ref{elip}) 
is of Kramers-Kronig type \cite{Landau}. 

There exists already significant progress in high-purity polarimetric techniques for  x-ray probes \cite{Marx2011,Marx2013}  which are  expected to  be exploited in the envisaged experiment at 
HIBEF  \cite{Schlenvoigt}. In first instance, the planned  polarimeter would be  designed to measure  vacuum  birefringence only. It  will  involve an analyzer set at a right angle to the initial  
polarization direction. This is justified  because, in a pure QED context with an x-ray probe and an optical strong field, the rotation of the incoming polarization plane is exponentially small 
[$\delta\vartheta_{\mathrm{QED}}=0$ for practical purposes]. In accordance, the transmission probability is   determined by the ellipticity induced by QED vacuum fluctuations 
$\mathcal{P}=\psi_{\mathrm{QED}}^2$ only. We remark that isolated  detections  of  both $\delta\vartheta(g,m)$ and $\vert\psi(g,m)\vert$ can  be carried out if  the analyzer  is  set  in such a way 
that the number of  counted photons  is  minimum. The axis of the analyzer would form an angle  $\frac{\pi}{2}+\delta\vartheta(g,m)$ with  respect to  the initial polarization plane whose measurement  
allows  for establishing  $\delta\vartheta(g,m)$. If the  
minimum count rate differs from zero, this would imply that the outgoing probe beam is elliptically polarized. Therefore, the polarization state transmitted by the analyzer reads 
$\pmb{\mathpzc{e}}=\pm\sin(\vartheta_0+\delta\vartheta)\pmb{\Lambda}_1\mp\cos(\vartheta_0+\delta\vartheta)\pmb{\Lambda}_2$. As a consequence, the transmission probability is 
$\mathcal{P}=\vert \pmb{\mathpzc{e}}\cdot \pmb{\varepsilon}\vert^2/\vert \varepsilon_{0}\vert^2=\left[\psi_{\mathrm{QED}}+\psi(g,m)\right]^2$. 
Assuming that $\psi_{\mathrm{QED}}>\psi(g,m)$, we find that the number of photons to be counted  reads 
\begin{equation}\label{flippingprobability}
\mathcal{N}=\mathcal{N}_{\mathrm{in}}\mathcal{N}_{\mathrm{shot}}\mathcal{T}\left[\psi_{\mathrm{QED}}^2+2\psi_{\mathrm{QED}}\psi(g,m)\right],
\end{equation}where $\mathcal{N}_{\mathrm{shot}}$  counts the number of laser shots used for a measurement,  $\mathcal{T}$ 
denotes  the transmission coefficient of  all optical components  and  $\mathcal{N}_{\rm in}$ is the  number of  probe  photons emitted in each shot. 
This expression shows that the number of signal  photon increases the greater  $\psi(g,m)$ is.

\section{Consequences of the pulse profile \label{secasymp}}

\begin{figure}
\centering
\includegraphics[width=7cm]{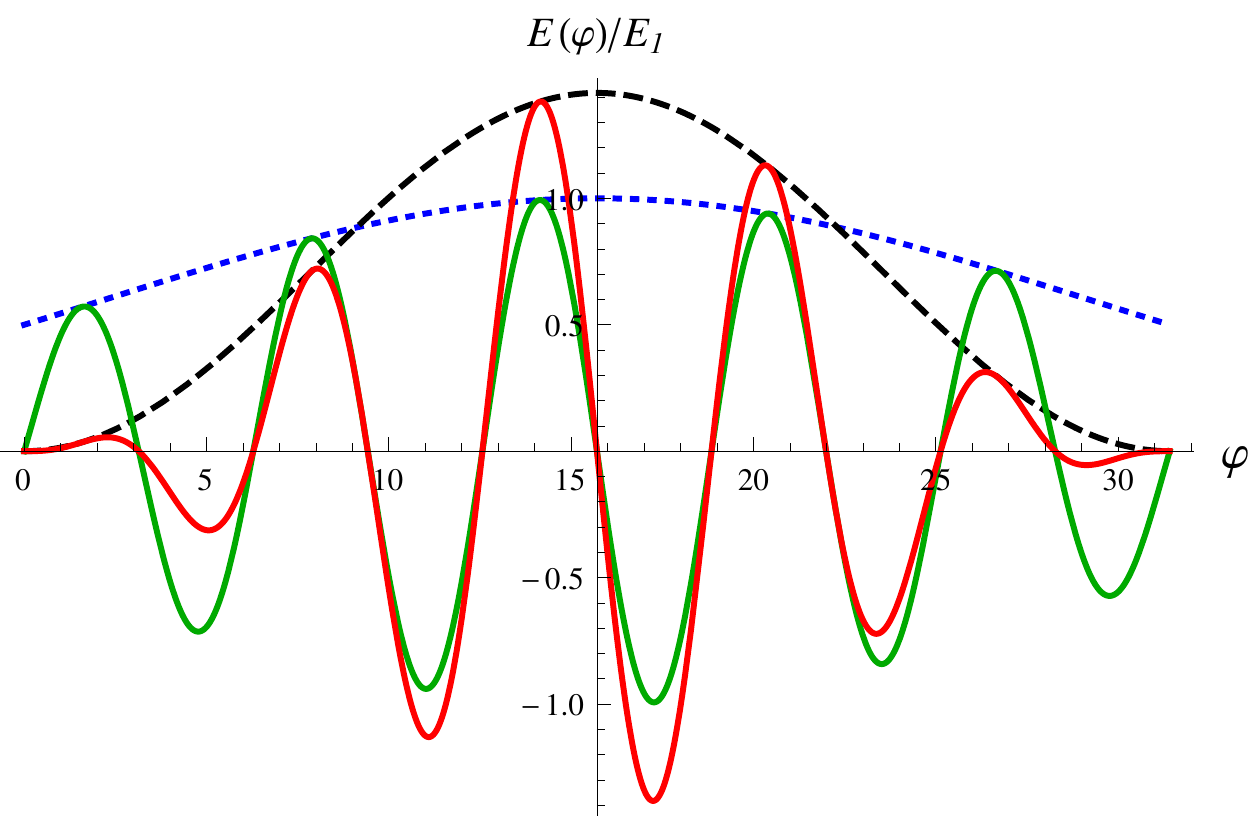}
\caption{\label{fig.002}Laser pulses with Gaussian and $\sin^2$-profiles are shown in green and red with $\mathpzc{N}=5$, respectively. While the pure Gaussian envelope is dotted in blue, the one associated with the 
$\sin^2$ function is dashed in black. These envelopes are given as references. As the $\sin^2-$pulse falls off faster than the Gaussian pulse, its  peak strength must be  higher ir order to contain the same total energy.}
\end{figure}

In this section we particularize  the optical obsevables given in  Eqs.~(\ref{rot}) and (\ref{elip}) for the cases in which the  external 
field is characterized  by a Gaussian or   a $\sin^2$ envelope [see Fig.~\ref{fig.002}].  Later, in sec.~\ref{genrat}, we generalize the 
results found for these two pulses by considering a generic pulse envelope and  a carrier envelope phase (CEP).

\subsection{Gaussian pulse}

We  wish  to investigate  the  optical  observables  induced by ALPs when  the  field profile function is   of the form
\begin{equation}
\psi_1^\prime(\varphi)=\exp\left[-\frac{(\varphi-\mathpzc{N}\pi)^2}{2\Delta\varphi^2}\right]\sin(\varphi).\label{fieldprofilefunction}
\end{equation}Here $\Delta\varphi=\pi\mathpzc{N}/\sqrt{2\ln(2)}$ is defined from  the full width at half maximum (FWHM) of the field  with $\mathpzc{N}$ 
referring to the number of oscillation cycles within  the Gaussian envelope.  The Fourier transform of this pulse  
allows us to express the rotation angle [see Eq.~(\ref{rot})] in the following form
\begin{equation}\label{rotsmalchi}
\begin{split}
&\delta\vartheta(g,m)\approx -\frac{1}{4}\sin(2\vartheta_0)\mathcal{P}_{\gamma\to\phi},\\
&\mathcal{P}_{\gamma\to\phi}=\frac{\pi}{2}\frac{g^2}{\varkappa_0^2}I \Delta\varphi^2e^{-\Delta\varphi^2(\mathpzc{n}_*^2+1)}\sinh^2\left(\Delta\varphi^2\mathpzc{n}_*\right).
\end{split}
\end{equation}
Assuming the condition $\Delta\varphi^2>\Delta\varphi^2 \mathpzc{n}_*\gg1$,  one can use the  approximation $\sinh^2(\Delta\varphi^2\mathpzc{n}_* )\approx\frac{1}{4}\exp[2\Delta\varphi^2\mathpzc{n}_*]$. 
Its substitution into  Eq.~(\ref{rotsmalchi})  leads  to  
\begin{eqnarray}
\begin{split}
&\delta\vartheta(g,m)\approx-\frac{1}{4}\sin(2\vartheta_0)\frac{\pi}{8}\frac{g^2}{\varkappa_0^2}I \Delta\varphi^2 e^{-\Delta\varphi^2(\mathpzc{n}_*-1)^2}.
\end{split}\label{Imdeltafinal}
\end{eqnarray}Since  the rotation angle  is maximized in the vicinity of $n_*=m^2/(2 \varkappa_+k_-)\approx1$,  we can anticipate that the most stringent bound  
will arise at  the  resonant  mass  $m_*\equiv\sqrt{2(\varkappa_+k_-)}$. On the other hand, since  $\Delta\varphi^2 n_*\ll1<\Delta\varphi^2$ implies  
$\sinh(\Delta\varphi^2\mathpzc{n}_*)\approx\Delta\varphi^2\mathpzc{n}_*$  [see  Eq.~(\ref{rotsmalchi})], we find that  $\vert\delta\vartheta(g,m)\vert\sim \mathpzc{n}_*^2\Delta\varphi^6\exp(-\Delta\varphi^2)$  
is  exponentially suppressed, which  indicates that in this regime  vacuum  dichroism  tends to vanish.
 
The ellipticity [see Eq.~(\ref{elip})] can be determined straightforwardly  by noting that the Hilbert transform of a Gaussian function 
$\frac{1}{\pi}\mathrm{P}\int_{-\infty}^{\infty}dz \frac{e^{-z^2}}{y-z}=\frac{2}{\sqrt{\pi}}\mathpzc{D}_F(y)$ is specified by  the Dawson integral $\mathpzc{D}_F(y)=e^{-y^2}\int_0^y dz\ e^{z^2}$  \cite{NIST}. 
With this detail in mind, we find 
\begin{equation}\label{Elipfinal}
\begin{split}
&\vert\psi(g,m)\vert\approx \frac{1}{2}\sin(2\vartheta_0)\frac{\sqrt{\pi}}{8}\frac{g^2 }{\varkappa_0^2}I \Delta\varphi^2\Big\vert\mathpzc{D}_F\left(\Delta\varphi\left[1+\mathpzc{n}_*\right]\right)\\&\qquad\quad\quad-\mathpzc{D}_F\left(\Delta\varphi\left[1-\mathpzc{n}_*\right]\right)-2 e^{-\Delta\varphi^2}\mathpzc{D}_F(\Delta\varphi \mathpzc{n}_*)\Big\vert.
\end{split}
\end{equation}Let us consider the asymptotic behavior of $\vert\psi(g,m)\vert$ as  $1\ll\mathpzc{n}_*$. Then,  
$\mathpzc{D}_F(\Delta\varphi(1\pm \mathpzc{n}_*))\approx\pm \mathpzc{D}_F(\Delta\varphi\mathpzc{n}_*)$,  and 
\begin{equation}
\begin{split}
&\vert\psi(g,m)\vert\approx \frac{1}{2}\sin(2\vartheta_0)\frac{\sqrt{\pi}}{4}\frac{g^2 }{\varkappa_0^2}I \Delta\varphi^2 \\ &\qquad\quad\ \times\left\vert \mathpzc{D}_F\left(\Delta\varphi\mathpzc{n}_*\right)\right\vert\left[1-e^{-\Delta\varphi^2}\right].
\end{split}
\end{equation}
If  $\Delta\varphi \mathpzc{n}_*\gg1$,  we can use   $\mathpzc{D}_F(y)\approx 1/(2y)-1/(4y^3)$ and find 
that $\vert\psi(g,m)\vert \approx\frac{1}{2}\sin(2\vartheta_0)\sqrt{\pi}g^2I \Delta\varphi/[8\varkappa_0^2\mathpzc{n}_*]$. 

Turning our attention to  the situation in which  $\mathpzc{n}_*\ll1$, the resulting  expression for the ellipticity  in this limit reads
\begin{equation}
\begin{split}
&\vert\psi(g,m)\vert\approx \frac{1}{2}\sin(2\vartheta_0)\frac{\sqrt{\pi}}{4}\frac{g^2 }{\varkappa_0^2}I \Delta\varphi^3\mathpzc{n}_* \Bigg\vert1-2\Delta\varphi\mathpzc{D}_F(\Delta\varphi)\\ &\qquad\qquad-e^{-\Delta\varphi^2}\frac{\mathpzc{D}_F\left(\Delta\varphi\mathpzc{n}_*\right)}{\Delta\varphi\mathpzc{n}_*}\Bigg\vert,
\end{split}\label{ndeltaphsmaller1}
\end{equation}where the defining equation of the Dawson function has been used,  $\mathpzc{D}_F^\prime(\Delta\varphi)=1-2\Delta\varphi\mathpzc{D}_F(\Delta\varphi)$. The last factor in Eq.~(\ref{ndeltaphsmaller1}), 
i.e. $\mathpzc{D}_F\left(\Delta\varphi\mathpzc{n}_*\right)/(\Delta\varphi\mathpzc{n}_*)\sim1$ as $\Delta\varphi\mathpzc{n}_*\ll1$. Conversely, the case $\Delta\varphi\mathpzc{n}_*\gg1$ can be   investigated through  
$\mathpzc{D}_F(\Delta\varphi)\approx\frac{1}{2\Delta\varphi}-\frac{1}{4\Delta\varphi^3}$, in which case the induced ellipticity   reads  $\vert\psi(g,m)\vert\approx\frac{1}{2}\sin(2\vartheta_0)\sqrt{\pi}g^2I \Delta\varphi\mathpzc{n}_*/[8\varkappa_0^2]$.   
The behavior of the ellipticity in a vicinity of the resonance $\mathpzc{n}_*\approx1$ is also of interest. In such a case, Eq.~(\ref{Elipfinal})  reduces to 
\begin{equation}\label{Elipfinalres}
\begin{split}
&\lim_{m\to m_*}\vert\psi(g,m)\vert\approx \frac{1}{2}\sin(2\vartheta_0)\frac{\sqrt{\pi}}{8}\frac{g^2 }{\varkappa_0^2}I \Delta\varphi^2\\&\qquad\qquad \qquad\times\Bigg\vert\mathpzc{D}_F\left(2\Delta\varphi\right)-2 e^{-\Delta\varphi^2}\mathpzc{D}_F(\Delta\varphi)\Bigg\vert.
\end{split}
\end{equation}For  $\Delta\varphi\gg1$, it leads to  $\vert\psi(g,m_*)\vert\approx\sin(2\vartheta_0)\sqrt{\pi}g^2I \Delta\varphi/[64\varkappa_0^2]$.


\subsection{$\sin^2-$pulse}

In order to evaluate the extent to which the projected bounds might depend on the pulse profile, we will now consider the case in which 
the  function $\psi_1^\prime(\varphi)$ is  of the form
\begin{equation}
\begin{split}
&\psi_1^\prime(\varphi)=\mathcal{R}\sin^2\left[\varphi/(2\mathpzc{N})\right]\sin(\varphi)\\
&=\frac{1}{2}\mathcal{R}\sin(\varphi)-\frac{1}{4}\mathcal{R}\sin(\lambda_+\varphi)-\frac{1}{4}\mathcal{R}\sin(\lambda_-\varphi)
\end{split}
\label{fieldprofilefunctionsin}
\end{equation}for  $\varphi\in[0,2\pi\mathpzc{N}]$ and zero otherwise. As before,  $\mathpzc{N}>1$ denotes the number of oscillation cycles within 
the $\sin^2-$envelope and  $\lambda_\pm=1\pm \mathpzc{N}^{-1}$.  The scaling parameter $\mathcal{R}^{2}\approx\frac{2}{3}\sqrt{\frac{2\pi}{\ln(2)}}$ is chosen in such a way that the total 
energy of the  pulse coincides with the one of the Gaussian pulse.

The use of the Fourier transform of Eq.~(\ref{fieldprofilefunctionsin})  allows us to express the rotation angle [see Eq.~(\ref{rot})] in the following form
\begin{equation}\label{rotsmalchisin}
\begin{split}
&\delta\vartheta(g,m)\approx -\frac{1}{4}\sin(2\vartheta_0)\frac{g^2}{4 \varkappa_0^2}\mathcal{R}^2I \sin^2\left(\pi  \mathpzc{n}_* \mathpzc{N}\right)\\&\qquad\qquad\qquad \times \left[\frac{1}{\mathpzc{n}_*^2-1}-\frac{\lambda_+}{2(\mathpzc{n}_*^2-\lambda_+^2)}-\frac{\lambda_-}{2(\mathpzc{n}_*^2-\lambda_-^2)}\right]^2.
\end{split}
\end{equation}This expression is characterized by three resonances: $\mathpzc{n}_*=1$ and  $\mathpzc{n}_*=\lambda_\pm$. While the former is already known from the analysis of the previous case 
[see below Eq.~(\ref{Imdeltafinal})],  the remaining ones define two additional resonant masses $m_{*\pm}=\sqrt{2\lambda_\pm\varkappa_+k_-}$ which  do not emerge  in the framework of the  Gaussian pulse. 
These extra resonances are direct consequences of the side-band terms arising in the spectral decomposition of the $\sin^2-$pulse, i.e. the  last two contributions 
in the second line in Eq.~(\ref{fieldprofilefunctionsin}). We note that the behavior of $\delta\vartheta(g,m)$ when $\pi\mathpzc{N}(\mathpzc{n}_*-1)\ll1$ or $\pi\mathpzc{N}(\mathpzc{n}_*-\lambda_\pm)\ll1$, i.e.,  
in a vicinity of $m_*$  and $m_{*\pm}$ is given, respectively, by 
\begin{equation}\label{rotsmalchisinlimits}
\begin{split}
\delta\vartheta(g,m)\approx-\frac{1}{4}\sin(2\vartheta_0)\frac{g^2\mathcal{R}^2I}{64 \varkappa_0^2} \pi^2 \mathpzc{N}^2\left\{
\begin{array}{c}\displaystyle
4\quad  \mathrm{for}\  m\approx m_*\\ \displaystyle
1\ \   \mathrm{for}\  m\approx m_{*\pm}
\end{array}\right.
\end{split}
\end{equation}Comparing the first line of this result with the outcome resulting from Eq.~(\ref{Imdeltafinal}) we find   that the projected 
sensitivity  expected from a $\sin^2-$pulse will be  smaller than the one corresponding to a Gaussian profile  by a factor $8/(9\mathcal{R}^2)\approx0.4$, approximately. Observe that 
Eq.~(\ref{rotsmalchisin})  tends to vanish  as  $\mathpzc{n}_*\gg \lambda_+ $ and $\mathpzc{n}_*\ll \lambda_-$. Hence, far from the resonant 
mass $m_*$ the projected bounds to be determined from the rotation angle  are expected to be less stringent.

Now  we focus on   the   ellipticity [see Eq.~(\ref{elip})]. In this case,  it is convenient to use  the   Hilbert transforms  
$\frac{1}{\pi}\mathrm{P}\int_{-\infty}^{\infty} \frac{dz\sin^2(z)}{(y-z)z^2}=\frac{1}{y}-\frac{\sin(2y)}{2y^2}$ and $\frac{1}{\pi}\mathrm{P}\int_{-\infty}^{\infty} \frac{dz\sin^2(z)}{(y-z)z}=-\frac{\sin(2y)}{2y}$
with which  we obtain 
\begin{equation}\label{Elipfinalsin}
\begin{split}
&\vert\psi(g,m)\vert\approx \frac{1}{2}\sin(2\vartheta_0)\frac{g^2\mathcal{R}^2I\pi\mathpzc{n}_*\mathpzc{N}}{16\varkappa_0^2}\left\vert\frac{1}{\mathpzc{n}_*^2-1}\right.\\&\quad+\frac{1}{4(\mathpzc{n}_*^2-\lambda_+^2)}+\frac{1}{4(\mathpzc{n}_*^2-\lambda_-^2)}-\frac{\sin\left(2\pi\mathpzc{n}_*\mathpzc{N}\right)}{\pi\mathpzc{n}_*\mathpzc{N}}\\&\qquad\qquad\times\left.\left[\frac{1}{\mathpzc{n}_*^2-1}-\frac{\lambda_+}{2(\mathpzc{n}_*^2-\lambda_+^2)}-\frac{\lambda_-}{2(\mathpzc{n}_*^2-\lambda_-^2)}\right]^2\right\vert.
\end{split}
\end{equation}Observe that $\vert\psi(g,m)\vert\to 0$ as  $\mathpzc{n}_*\gg \lambda_+ $ and $\mathpzc{n}_*\ll \lambda_-$, whereas for $\mathpzc{n}_*-1\ll(2\pi\mathpzc{N})^{-1}$ we obtain
\begin{equation}\label{Elipfinalsin2}
\begin{split}
&\vert\psi(g,m)\vert\approx \frac{1}{2}\sin(2\vartheta_0)\frac{g^2\mathcal{R}^2I\pi\mathpzc{N}}{64\varkappa_0^2}\left\vert\frac{4\lambda_++1}{1-\lambda_+^2}+\frac{4\lambda_-+1}{1-\lambda_-^2}\right\vert.
\end{split}
\end{equation}We point out that, for  $\mathpzc{N}\gg1$,   the expression above coincides with the corresponding  outcome resulting from a Gaussian pulse [see below Eq.~(\ref{Elipfinalres})].

\subsection{Generalization:  $f(\varphi)$-pulse \label{genrat}}

The laser pulses discussed in the previous sections can be understood  as particular cases  of a more general situation  in which the profile  function is given by 
\begin{equation}
\psi_1^\prime(\varphi)=\mathcal{R}f(\varphi)\cos\left(\varphi+\varphi_{\rm CEP}\right).
\end{equation}Here, $f(\varphi)$ with $f(\varphi)>0$ and  $f(\pm\infty)=0$ is a real analytic function in the upper half plane which is maximized at $\varphi=\pi\mathpzc{N}$, whereas  $\varphi_{\rm CEP}$ is  
the  CEP. As before, the scaling parameter:
\begin{equation}
\mathcal{R}^{2}=\frac{\sqrt{\pi}\Delta\varphi\left(1-e^{-\Delta\varphi^2}\cos(\pi\mathpzc{N})\right)}{2\int_{-\infty}^\infty d x^3 f^2(\varphi)\cos^2(\varphi+\varphi_{\rm CEP})},
\end{equation} guarantees that the pulse energy is invariant with respect to the chosen profile. Moreover, it must be understood that $\mathcal{R}^{2}$ does not depend on time 
and an explicit evaluation of it can be done by setting $x^0=0$.
In this context, the angle rotated by the polarization plane  [see Eq.~(\ref{rot})] 
\begin{eqnarray}\label{rotCEP}
\begin{split}
&\delta\vartheta(g,m)\approx-\frac{1}{4}\sin(2\vartheta_0)\mathcal{P}_{\gamma\to\phi}(\infty),\\ 
&\mathcal{P}_{\gamma\to\phi}(\infty)=\frac{g^2I\mathcal{R}^{2}}{16\varkappa_0^2}\vert\tilde{f}(\mathpzc{n}_*+1)+\tilde{f}(\mathpzc{n}_*-1)e^{-2i\varphi_{\rm CEP}}\vert^2
\end{split}
\end{eqnarray} coincides with an  outcome  obtained in Ref.~\cite{Villalba-Chavez:2013bda} upto the scaling factor $\mathcal{R}^{2}$. There,  this formula  was established  by computing 
$\mathcal{P}_{\gamma\to\phi}(\infty)$ via the S-matrix element associated with the  photon-ALP oscillations and by exploiting the relation between this quantity and the absorption coefficient  
[see discussion above Eq.~(\ref{rot})]. The equivalence between both procedures is  expected because  the optical theorem establishes that   $\mathcal{P}_{\gamma\to\phi}(\infty)$ is determined 
by  the imaginary part of the vacuum polarization tensor depicted in Fig.~\ref{fig.001}.

We should however indicate that, in the aforementioned reference,  the corresponding expression for the ellipticity  was not derived. According to  Eq.~(\ref{elip}),  this observable  reads 
\begin{eqnarray}\label{elipCEP}
\begin{split}
&\vert\psi(g,m)\vert\approx \frac{1}{2}\sin(2\vartheta_0)\frac{g^2I\mathcal{R}^{2}}{32\varkappa_0^2}\left\vert-\frac{1}{\pi}\mathrm{P}\int_{-\infty}^{\infty} \frac{d\eta}{\eta-\mathpzc{n}_*}\right.\\ &\quad\qquad\qquad\qquad\qquad\times\left.\vert\tilde{f}(\eta+1)+\tilde{f}(\eta-1)e^{-2i\varphi_{\rm CEP}}\vert^2\frac{}{}\right\vert.
\end{split}
\end{eqnarray}In these formulae, $\tilde{f}(\alpha)=\int d\varphi f(\varphi)e^{i\alpha\varphi}$ is the Fourier transform of the shape function $f(\varphi)$. The presence of $\tilde{f}(\mathpzc{n}_*\pm 1)$ 
in  Eqs.~(\ref{rotCEP}) and (\ref{elipCEP}) manifests that the  projected exclusion regions will  depend  on the envelope function of the external laser pulse. Besides, these formulae  indicate that the CEP allows 
for  interference between  the   $\tilde{f}(\mathpzc{n}_*+ 1)$ and $\tilde{f}(\mathpzc{n}_*- 1)$ terms whenever $\varphi_{\mathrm{CEP}}\neq(2\mathpzc{k}+1)\pi/2$ and $\mathpzc{k}\in\mathbb{Z}$. This interference 
effect  may  be constructive or destructive, depending on  the overall sign of the associated contribution.  Therefore, an appropriate choice of the CEP  may help to  optimize the  optical  signals.  
For the pulses  analyzed previously [see Eqs.~(\ref{fieldprofilefunction}) and (\ref{fieldprofilefunctionsin})],  this  occurs for $\varphi_{\mathrm{CEP}}=2\mathpzc{K}\pi$ with $\mathpzc{K}\in\mathbb{Z}$. 

\section{Experimental prospects}

First  we  estimate the projected  limits  considering the benchmark parameters of the  proposed experiment at  HIBEF \cite{Schlenvoigt}. In this setup the strong field  will be produced by  a  Petawatt 
laser  operating in the optical  regime with  $\varkappa_0\approx1.55\ \rm eV$ [$\lambda_0=800\ \rm  nm$], a repetition rate  of $1\ \rm Hz$, a temporal pulse length of about $30 \ \rm fs$ [$\Delta\varphi\approx11\pi$], 
and  a peak intensity $I\approx 2\times10^{22}\ \rm  W/cm^2$. The envisaged probe beam is  the European  x-ray free electron laser, operating with frequency $\omega_{\pmb{k}}=12.9\ \mathrm{keV}$ and delivering 
$\mathcal{N}_{\mathrm{in}}\approx 5\times 10^{12}$ photons per shot. The  transmission coefficient of the optics to be used in this experiment is  $\mathcal{T}=0.0365$, and the incoming polarization angle will be  
$\vartheta_0=\pi/4$. Under such a conditions,  a QED signal as small as $\vert\psi_{\mathrm{QED}}\vert=(9.8\pm6.7)\times10^{-7}\ \rm rad$ is  likely to  be reached provided  a perfect overlapping between the probe and 
the strong laser field is achieved \cite{Schlenvoigt}. 

\begin{figure*}[t]
\centering
\includegraphics[width=9cm]{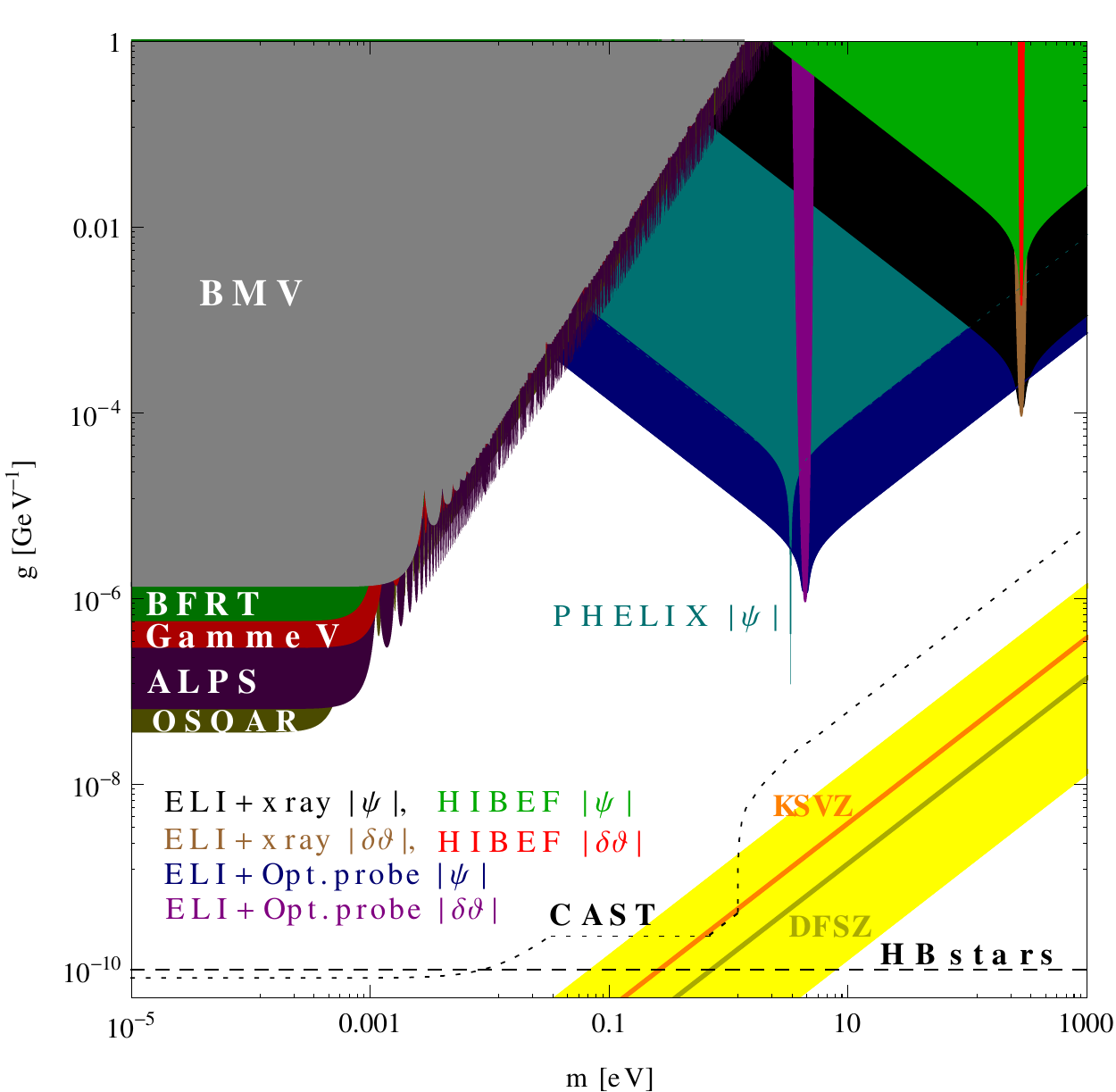}
\includegraphics[width=9cm]{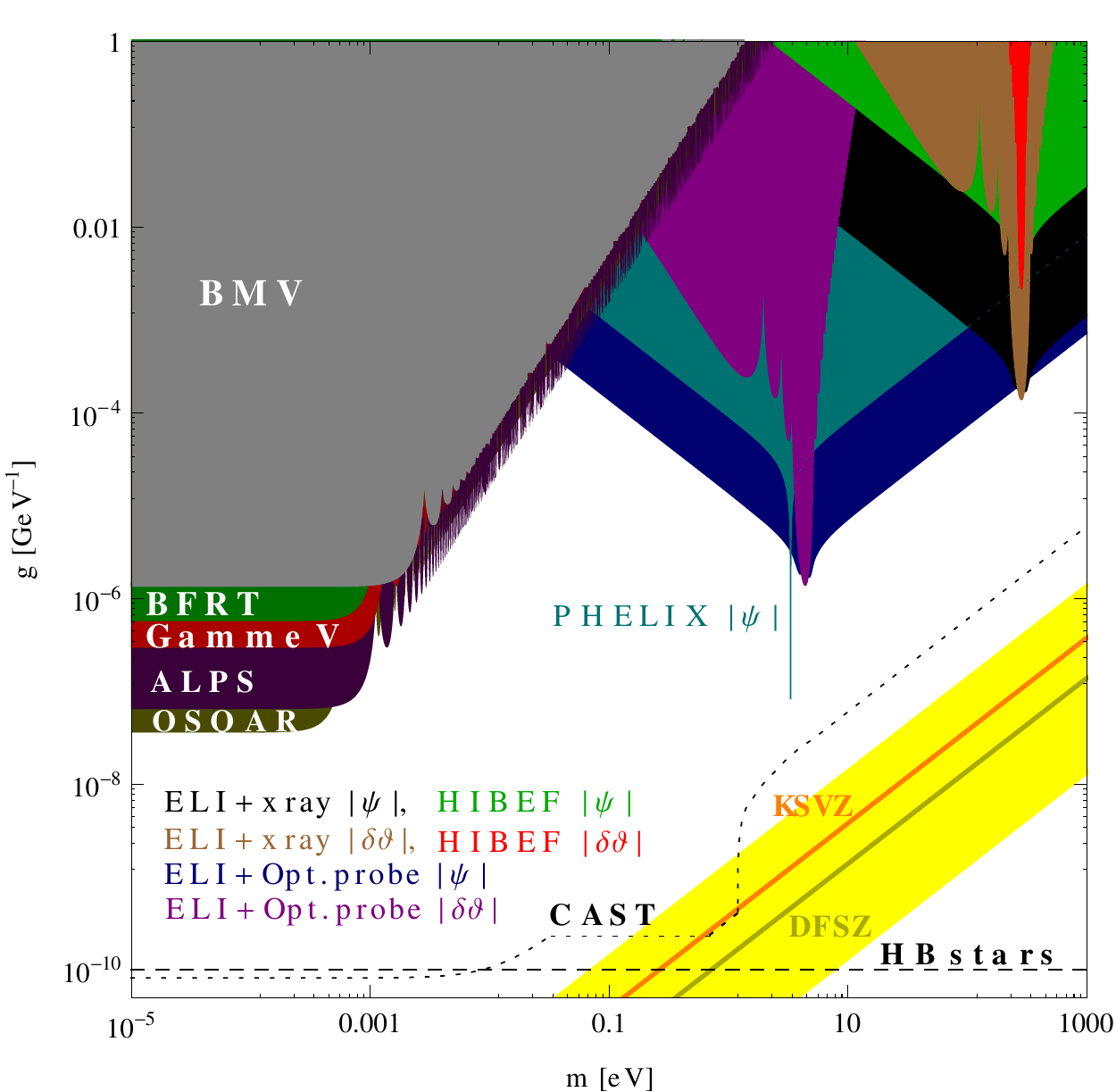}
\caption{\label{fig:mb003} Exclusion regions in the $(g,m)$-plane obtained from a polarimetric setup driven by  an intense linearly polarized laser pulse. The left (right) panel 
depicts the results based on  the Gaussian ($\sin^2$) pulse model.
Here the green, black, cyan and blue  shaded areas were determined from the ellipticity  induced by  ALPs on the initial polarization plane Eq.~(\ref{elip}), the brown, purple  and red  
wedges were found from the rotation Eq.~(\ref{rot}). The respective resonant peaks occur at $m_*=4.4\ \rm eV$ [purple] and $m_*=282.8\ \rm eV$ [brown, red].
The inclined yellow band covers the predictions of the axion models with $\vert E/N-1.95\vert=0.07-7$ [the notation of this formula is in accordance with Ref. \cite{CAST2011}]. The 
constraints resulting from HB stars [dashed line] are also shown. Further exclusion regions [shaded areas in the upper left corner] provided by  different LSW
experiment have been included too [see legend]. The exclusion limit resulting from  solar monitoring of a plausible ALP flux [CAST experiment] has been included as well [dotted line]. 
We point out that the upper bound resulting from such an experiment strongly oscillates in the mass region  $0.4 \ \mathrm{eV}\leqslant m\leqslant0.6\ \rm eV$. This oscillatory pattern 
has been replaced by a straight dotted line, corresponding to the exclusion limit $g\leqslant2.3\times 10^{-10}\ \rm GeV^{-1}$, established in \cite{CAST2011} at $95\% $ confidence level. 
For the exact picture of the CAST exclusion limits, we refer the reader  to the original publication \cite{CAST2011,CAST2014}. 
 }
\end{figure*}

An exclusion region can be inferred from this projected result  by assuming that the induced ellipticity due to ALPs does not overpass   the upper bound set by $\vert\psi_{\mathrm{QED}}\vert$. It  is  
shaded in green in the right upper corners in Fig.~\ref{fig:mb003}. While the outcome  shown in  the left panel relies on the Gaussian model, the one  in the right panel is based on the $\sin^2-$pulse. 
In each panel, there is a tiny  wedge shaded in red,  which is  ruled out  by supposing that the rotation angle  can be  measured with a sensitivity of the same order of magnitude 
of $\vert\psi_{\mathrm{QED}}\vert$.  Our estimates reveal  that the most  stringent bounds  $g<1.4\times 10^{-3}\ \rm GeV^{-1}$ [Gaussian profile] and $g<2.2\times 10^{-3}\ \rm GeV^{-1}$ 
[$\sin^2-$profile] would  emerge  at the resonant mass $m_*\approx282.8\ \rm eV$ [see below Eq.~(\ref{Imdeltafinal})]. 

As we already pointed out, the energy scale associated with the waist size of the pulse  $w_0$ limits the validity of our predictions towards smaller ALP masses.  
In view of the strong focusing applied at HIBEF [$w_0\approx 2\lambda_0$], our potential discovery applies for  $m\gg 0.12\ \rm eV$.  We must, in addition, mention   that this result relies on 
the forward scattering  analysis,  and that the   waist size of the probe  $w_{\rm probe}=42.5\ \rm\mu m$ is bigger than  $w_0$. This situation in combination with the nonconservation of the  transverse momentum 
that the focusing induces,  is favourable  to  scatter  probe photons slightly off  the forward direction. Such an effect has been proposed as an alternative way to detect the QED birefringence  at a small angle 
\cite{Karbstein:2015xra}. It   might  as well be  beneficial for the search of ALPs as the  signal-to-noise ratio for photons transmitted through the analyzer improves notably.  However, this study would 
require to incorporate the focusing effects, which is   still  beyond the scope of this work.

If the planned   HIBEF experiment  was driven by the strong field to be reached  at ELI  [$I\approx 10^{25}\ \rm W/cm^2$, $\varkappa_0\approx1.55\ \rm eV$, $\tau\approx 13 \ \rm fs$, 
corresponding to $\Delta\varphi\approx4\pi$],  and the sensitivity   remained  within the same order of magnitude  $\sim 10^{-6}$,  the limits above would be  pushed down to  $g<9.3\times 10^{-5} \ \rm GeV^{-1}$ 
[Gaussian model]  and $g<1.4\times 10^{-4} \ \rm GeV^{-1}$ [$\sin^2-$pulse]  at the resonant  mass $m_*\approx282.8\ \rm eV$.  The projected areas to be excluded from the  ellipticity 
[rotation angle]  can be seen in Fig.~\ref{fig:mb003} in black [brown]. We emphasize that  the pulse at ELI is  expected to be strongly focused [$w_0\sim \lambda_0$]. Hence,  the  exclusion areas found 
for this setup  are expected to be trustworthy   as long as $m\gg 0.24 \ \rm eV$. Our  estimates reveal  that the shape of the bounds resulting from the ellipticity  almost coincide for 
both pulse models. However,  the  borders of the excluded areas  coming from the rotation angle  differ from each other more strongly.

The described behavior  is  even more pronounced  when both observables  are probed with an optical laser beam and the envisaged ELI laser drives  the vacuum polarization. The potential exclusion regions  
associated with this case are summarized in Fig.~\ref{fig:mb003} in blue [ellipticity] and purple [rotation angle].  They have been found by supposing  a sensitivity of the order of $\sim10^{-10}\ \mathrm{rad}$,  
a probe frequency  $\omega_{\pmb{k}}=2\varkappa_0=3.1\ \rm eV$ and a  probe   intensity  much smaller than the one of  the strong laser field. We remark that the outcomes resulting from this hypothetical setup 
were obtained  by considering  a counterpropagating geometry $\varkappa_+k_-=2\varkappa_0\omega_{\pmb{k}}$ and an initial polarization angle $\vartheta_0=\pi/4$. 

To conclude, we study a  situation in which the  strong field is generated by  the nanosecond front-end of the PHELIX laser \cite{PHELIX}, [$I\approx 10^{16}\  \rm W/cm^2$, $w_0\approx 100-150 \ \mu \mathrm{m}$,
$\varkappa_0\approx1.17\ \rm eV$, $\tau\approx20 \ \rm ns$, corresponding to  $\Delta\varphi\approx 5\times 10^6 \pi$]. It is worth mentionig that the electromagnetic pulse  produced by this system closely approaches to 
a monochromatic plane wave as the conditions $\tau\gg \varkappa_0^{-1}$  and $w_0\gg\lambda_0$  are fulfilled simultaneously. The potential exclusion bounds which follow from this setup are  valid whenever the condition 
$m\gg w_0^{-1}\approx1.3 \ \rm meV$ is satisfied.  As in the previous  case, we take the probe beam with a doubled frequency 
$\omega_{\pmb{k}}=2\varkappa_0=2.34\ \rm eV$. Besides, we  suppose that its intensity and  waist size are much smaller than the corresponding quantities of the strong field. Assuming some achievable  conditions 
such as  $\vartheta_0=\pi/4$, a counter propagating geometry and  a sensitivity  $\vert\psi(g,m)\vert \lesssim 10^{-10}\ \mathrm{rad}$, we find  that the area 
shaded in cyan [Fig.~\ref{fig:mb003}] could be excluded potentially. This projected result shows that large  sensitivities can be achieved,   provided the number of cycles $\mathpzc{N}$ is large enough   to compensate  
for the relative smallness of the laser intensity $I$. 

\section{Conclusions}

We have studied the discovery potential  that  modern and envisaged laser systems  offer in the search for ALPs.
Our investigation reveals that laser-based setups, designed to detect the  hitherto unobserved QED vacuum birefringence, may provide  stringent 
bounds on the ALP-diphoton coupling $g$  in mass regions where  the constraints resulting from experiments driven by di\-pole magnets 
are considerably less severe.  Special attention has been paid to the consequences resulting from  the pulse profile function, 
which were evaluated  by solving  perturbatively  the system of equations describing the oscillation of photons into ALPs mediated by 
a generic plane-wave background. Our analysis  points out that  a broad   sector  of the parameter space of ALPs might be discarded, 
no matter   what  type of  strong field profile is utilized. The precise location of the projected exclusion areas  
will depend, in general, on the chosen  pulse profile and the CEP. However, a direct comparison  between two pulses with different 
envelope has indicated that the outcomes resulting from the ellipticity are less sensitive to this  dependence  than the bounds arising 
from a plausible  rotation  of the polarization plane.  Besides, we have pointed out that an appropriate choice of the CEP  may  optimize 
the ALPs search.

\vspace{0.005 in}
\begin{flushleft}
\textbf{Acknowledgments}
\end{flushleft}
\vspace{0.005 in}
S. Villalba-Ch\'avez and C. M\"{u}ller gratefully acknowledge the funding by the German Research Foundation (DFG) under Grant No. MU 3149/2-1.

\end{document}